# Resource Allocation in Public Cluster with Extended Optimization Algorithm

Z. Akbar and L.T. Handoko
Group for Theoretical and Computational Physics, Research Center for Physics, Indonesian Institute of Sciences,
Kompleks Puspiptek Serpong, Tangerang 15310, Indonesia

*Abstract-* We introduce an optimization algorithm for resource allocation in the LIPI Public Cluster to optimize its usage according to incoming requests from users. The tool is an extended and modified genetic algorithm developed to match specific natures of public cluster. We present a detail analysis of optimization, and compare the results with the exact. We show that it would be very useful and could realize an automatic decision making system for public clusters.

## I. INTRODUCTION

Since 2005, we have maintained LIPI Public Cluster (LPC) [1]. LPC is a kind of cluster computing that is open for public for completely free [2,3]. Due to its natures, LPC implements a unique architecture, so called Multi Block Approach to enable simultaneous parallel computings in different blocks of cluster by several anonymous users [4].

In LPC, the users have a freedom to upload their own parallel programmings and to run them on it once their requests have been approved and the blocks of cluster have been assigned. A block of cluster could consist of several, normally ranging between 2 till 5 nodes. The approved number of nodes is determined manually by the administrator according to the request, the scale of parallel programming and of course the availability at the time of registration. All provided blocks are activated immediately after reviewing and approval procedures. Since LPC consists of many nodes with varying specifications and calculation powers, the main concern in assigning a block is total efficiency and performance.

This is completely different with any conventional clusters existing around the world. A conventional cluster is normally assigned for a particular group in certain period. The whole nodes are then designed to work together without any interruption by another groups. Moreover, the computational jobs sent to the cluster should be developed with strictly considering the characteristics of cluster to obtain desired performances.

In contrast, users in LPC are anonymous and at the time of registration they do not acknowledge really available nodes nor which nodes will be provided. This is crucial since in parallel programmings the users divide the whole jobs into smaller tasks that in most cases are not homogeneous, i.e. they must distribute the tasks properly to match the available nodes in order to obtain optimum performance. This means LPC, which has many nodes with various specifications and then many combinations of them, should have enough flexibilities to accommodate various parallel programmings and requests from users.

Therefore, we put a regulation for every user to propose preferred power ratios of each node for the whole nodes they request regarding the characteristics of parallel programmings being run on LPC. After that the administrator seek for the best combination of available nodes being provided to users.

During the running period since its initial launching, we have faced problems on how to allocate the resources (nodes) properly according to the incoming requests. This problem was not so seriously taken previously. However, since the number of nodes is increasing till 45 at current condition, the resource allocation is getting complicated and non-trivial. Then we have developed a tool to accomplish this task. However we should emphasize that the terminology of resource allocation is slightly different with the conventional one. Resource allocation in conventional sense is associated with the way to distribute available resources in all nodes to improve computational performance during its running period. In our case, it means resource allocation before the running period to obtain the best combination from available nodes according to the user request. Actually in LPC resource allocation during running period is irrelevant, since a user sticks at the pre-allocated block. Of course, it is still relevant in the sense of resource allocation among nodes inside a block.

Now we are going to discuss the algorithm we have developed and implemented to overcome the problem mentioned above. First we discuss a general prescription for optimization in LPC, followed by the modified genetic algorithm for the selection rule. Before concluding, we present detail analysis to show how the algorithm works and compare it with the exact calculation.

## II. OPTIMIZATION ALGORITHM : FITNESS

As briefly mentioned above, in allocating resources we have to always deal with the optimization. The reason is simply because the number of possible combinations are in most cases astronomically huge. For example, to choose 3 nodes for a certain user among 20 available nodes one has $20!/[(20 - 3)! \times 3!]$ combinations, since we do not distinguish its orders. The number is getting much larger for even only one order more available nodes or number of nodes requested.

TABLE I
The ratios of node's capacites acording to its specifications.

| Node | Capacity (%) | Category |
|---|---|---|
| 1 | 80 | II |
| 2 | 90 | II |
| 3 | 100 | I |
| 4 | 100 | I |
| 5 | 90 | II |
| … | … | … |

TABLE 2
An illustration on the total performances and deviations of 3 requested nodes (in %).

| Solution | Node-1 | Node-2 | Node-3 | Total |
|---|---|---|---|---|
| [1,2,3] | 80 | 90 | 100 | 270 |
| [4,5,7] | 100 | 90 | 50 | 240 |
| [1,10,9] | 80 | 70 | 50 | 200 |
| … | … | … | … | … |

The optimized combination is determined by particular conditions, so called fitness, that are highly depending on the case under consideration [5]. In LPC, we have define the following procedures and conditions for fitness :
1. Since the node specifications in LPC are heterogen, we predefine each node with a percentage indicating its computational capacity against the node with highest specification.
2. Further all nodes with same percentage, that is having comparable computational capacities, are classified as the same class.
3. Number of node allocation and its computational capacities are based on the user request during the initial registration.

The first and second conditions are illustrated in Tab 1. In the table, node 3 has the highest overall computational power (concerning its memory, processor, space, connection speed, etc). The remaining nodes, with lower specifications, are then assigned with appropriate percentages in the scale of 0-100%. Further, we classify all nodes to some classes where each class consists of nodes with comparable computational powers. Through the paper this ratio of capacity is denoted as $R_{cap\text{-}node}$.

The third condition is quite different with regular clusters. In regular cluster allocating nodes is done automatically during running process according to (physically) available nodes in the cluster, because in principle present users are provided with the whole cluster. In contrast, in LPC the whole cluster is divided into several blocks that might be used by several (different) users simultaneously. So, by this nature at the time of registration new users are asked to request the number of nodes ($N_{request}$) and the ratio of capacities needed for each child process ($R_{cap\text{-}child}$) in their parallel programmings in the scale of 0-100%. This should be determined by the users concerning the characteristics of their own parallel programmings. Just to mention, in contrary in regular cluster the users should consider the characteristic of cluster in developing the parallel programmings. The fitness and then the resource allocation is done automatically to sustain the best performance during running time although, for instance, some nodes are getting down, etc.

Now we are ready to formulate the above fitness conditions. First of all, any combinations of assigned nodes with the number as requested by users should by definition fulfill :

$$\sum_{n=1}^{Nrequest} R_{cap-node}^{(n+1)} \cong 100\% \pm T \qquad (1)$$

where $T$ means the tolerance from the ideal desired percentage, i.e. 100%. Later on we take $T$ to be 5% for a moderate number in the case of LPC.

We must also find the best combination as desired by users to match the ideal ratios of child processes in their parallel programmings. We have defined the total deviations between two neighboring nodes to determine its fitness. This is formulated by the following equation :

$$\sum_{n=1}^{Nrequest} \left( R_{cap\_child}^{(n+1)} - R_{cap-child}^{(n)} \right) \cong$$
$$\sum_{n=1}^{Nrequest} \left( R_{cap-node}^{(n+1)} - R_{cap-node}^{(n)} \right) \pm \qquad (2)$$
$$\left( T \times \left( \sum_{n=1}^{Nrequest} R_{cap-child}^{(n+1)} - R_{cap-child}^{(n)} \right) \right)$$

which can then be rewritten as :

$$\sum_{n=1}^{Nrequest} \left( R_{cap-node}^{(n+1)} - R_{cap-node}^{(n)} \right) \cong$$
$$(1 \pm T) \times \left( \sum_{n=1}^{Nrequest} \left( R_{cap-child}^{(n+1)} - R_{cap-child}^{(n)} \right) \right) \qquad (3)$$

We further deploy Eqs. (1) and (3) as the master equations to determine the total fitness of a combination. As an illustration, let us consider a user who has requested 3 nodes with the ratio of child processes is 50% : 30% : 20%, while we have 10 nodes available at the moment with ratio of capacities : $R_{cap\text{-}node}(1) = 80\%$, $R_{cap\text{-}node}(2) = 90\%$, $R_{cap\text{-}node}(3) = 100\%$, $R_{cap\text{-}node}(4) = 100\%$, $R_{cap\text{-}node}(5) = 90\%$, $R_{cap\text{-}node}(6) = 90\%$, $R_{cap\text{-}node}(7) = 50\%$, $R_{cap\text{-}node}(8) = 80\%$, $R_{cap\text{-}node}(9) = 50\%$ and $R_{cap\text{-}node}(10) = 70\%$. Since the order of nodes in a combination is irrelevant, the total considerable combinations is $10!/[(10 - 3)! \times 3!] = 120$. Some combinations are given in Tab. 2 showing the total $R_{cap\text{-}node}$ and its deviations. Best values indicate the requested values which should be achieved.

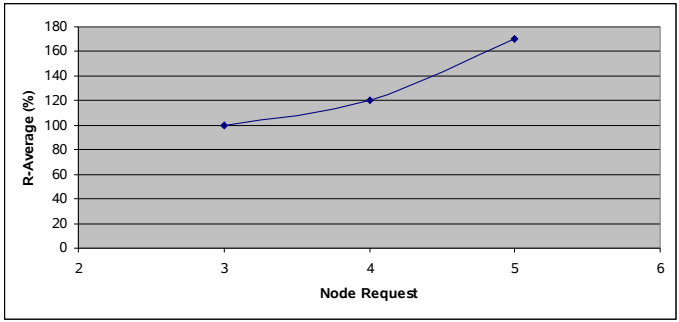

Fig. 1. The exact calculation of the average of R for totally 20 available nodes in the case of 3, 4 and 5 requested nodes.

So far, it is clear that this rule could lead to an astronomical number of combinations for more available and / or requested nodes. For practical reason in LPC where the approval and assignment procedures should be done online in as short time as possible, this algorithm lacks of convenience. Then we continue using this fitness and together deploying the genetic algorithm to reduce the number of combinations significantly [6].

III. SELECTION RULE : EXTENDED GENETIC ALGORITHM

Genetic Algorithm (GA) is a method widely used for optimization problem inspired by principle of evolution in life science [7]. GA is based on the idea that the optimization problem can be seen as an "individual" with particular characteristics coded by a set of finite parameters. These parameters could be genes which form a chromosome representing individual structure in a real world, in our case is the solution of optimization problem. These individuals further create generations which form a population to find a fittest individual. Only the fittest individuals will survive through next generations. This procedure is repeated for some generations till we find the optimum value according to particular conditions as the result.

GA has been applied on cluster as load balancing and task allocation [7,8,9]. In this paper we apply this algorithm but with some modifications to fit unique characteristics of LPC. Further we call it as the Extended Genetic Algorithm (EGA). In analogy with the conventional GA, The individual and genes are analogous to the combination of nodes and nodes itself.

In LPC, we have deployed the following procedures inspired by GA :

1. Listing all available nodes at the moment with its defined number and $R_{cap-node}$'s.
2. Generating considerable combinations randomly according to the user request, i.e. $N_{request}$ and $R_{cap-child}$ under conditions :
   - Each combination must not contain multiple nodes with same numbers.
   - The order of nodes in a combination is irrelevant.
   - For assumption, the number of randomly generated combinations should equal or larger than the number of available nodes. A set of these generated combinations is then as the initial population.
3. Operating GA in all generated combinations in the above initial population to obtain new combinations using :
   - Cross-over method :
     Nodes inside two different combinations which have been choosed randomly are interchanged. Following the above condition, some combinations containing same nodes are thrown away.
   - Swapping mutation method [10] :
     A combination is choosed randomly, and two nodes inside it are interchanged. According to the above condition, this would not generate new combination, and only results same fitness which increases its surviving probability to the next generation.
4. The fitness of each newly generated combinations are calculated using Eq. (3).
5. Selecting the appropriate combinations which survive to the next generation using roulette-wheel method. All obtained fitness are rescaled in the scale of 0-100% and are mapped in a circle with the angles associated to its percentages. Using the stochastic sampling with replacement [6], a new combination is randomly retrieved.
   In LPC, in order to guarantee that the best combination in each generation is survived, we pass it to the next generation, while only the remaining combinations are selected using roulette-wheel method.
6. All selected combinations form a new generation, and the same procedures in no. 2 till 5 are repeated till we obtain the optimum (closest) values as requested by users.

We have completed all methods and procedure in our EGA for the optimization algorithm deployed in LPC. In the subsequent section we present a simulation for a typical case in LPC.

IV. RESULT AND ANALYSIS

Now we simulate EGA in a particular case of 20 available nodes with various numbers of requested nodes and its $R_{cap-child}$'s : 1) $N_{request}$ = 3 (50% : 30% : 20%), 2)

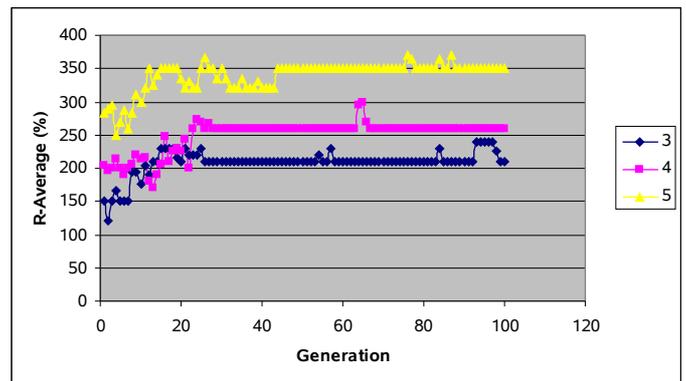

Fig. 2. The average of generated R vs number of generations using EGA for totally 20 available nodes in the case of 3, 4 and 5 requested nodes with 20 combinations in a population.

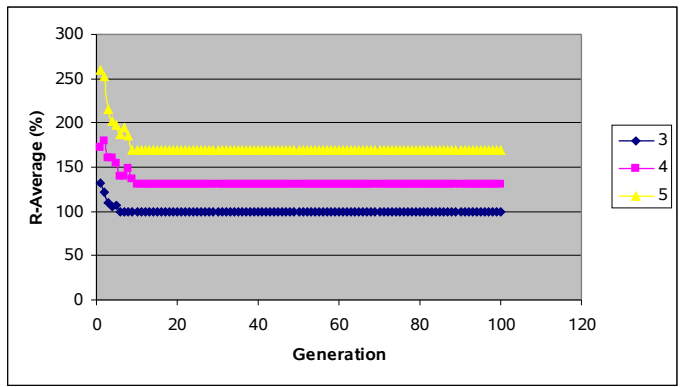

Fig. 3. The average of generated R vs number of generations using EGA for totally 20 available nodes in the case of 3, 4 and 5 requested nodes with 40 combinations in a population.

$N_{request}$ = 4 (40% : 20% : 20% : 20%) and 3) $N_{request}$ = 5 (30% : 20% : 20% : 20% : 10%).

As assumed in the preceding section, the number of randomly generated combinations should equal or larger than the number of available nodes. We choose the typical values that is 20 and 40 combinations.

First of all we perform an exact calculation for all of them. The result is shown in Fig. 1. We can deduce from the figure that for $N_{request}$ = 3 we can obtain almost an ideal combination, i.e. 100% matches the user request. On the other hand, $N_{request}$ = 5 case results 170% efficiency. We note that lower values than 1 means the assigned combination of nodes is under-spec that leads to lower performance of overall computation. While larger values than 1 indicates over-spec which means wasting the resources.

Now we simulate the calculation using EGA. In Figs. 2 and 3 we have plotted the same as above for 3 cases with number of combinations is 20 and 40. It is straightforward to deduce that increasing the number of combinations in a population would fasten reaching the optimum values, that is the values are close to the exact calculations as shown in Fig. 1. In the case of 20 combinations the optimum value is reached after 40 generations, while for 40 combinations it happens after 15 generations.

## V. CONCLUSION

Finally we have proposed an alternative algorithm for optimization problem that is more appropriate for the case of public cluster like LPC. The algorithm has proven good and reliable performances for the desired purposes on assigning an appropriate combination of nodes according to the incoming request of anonymous users.

Finally we would like to mention that the EGA is going to be embedded in the web-interface of LPC as an integrated and automatic decision making system to enable instant approval and assignment process of the best combination of nodes for our users [11].

## ACKNOWLEDGMENT

This work is financially supported by the Riset Kompetitif LIPI in fiscal year 2007 under Contract no. 11.04/SK/KPPI/II/2007 and the Indonesia Toray Science Foundation Research Grant 2007.

## REFERENCES

[1] LIPI Public Cluster, *http://www.cluster.lipi.go.id*.
[2] L.T. Handoko,, Indonesian Copyright, No. B 268487 (2006).
[3] Z. Akbar, Slamet, B.I. Ajinagoro, G.J. Ohara, I. Firmansyah, B. Hermanto and L.T. Handoko, "Open and Free Cluster for Public", Proceeding of the International Conference on Rural Information and Communication Technology 2007, Bandung, Indonesia,2007.
[4] Z. Akbar, Slamet, B.I. Ajinagoro, G.J. Ohara, I. Firmansyah, B. Hermanto and L.T. Handoko, "Public Cluster : parallel machine with multi-block approach", Proceeding of the International Conference on Electrical Engineering and Informatics, Bandung, Indonesia, 2007.
[5] C. H. Papadimitriou and K. Steiglitz, Combinatorial Optimization: Algorithms and Complexity, Dover Publication, 1998.
[6] For example see :
D. Whitley, "A Genetic Algorithm Tutorial", Computer Science Department, Colorado State University, unpublished.
[7] Foncesa, C.M. and Fleming, C.J., "Genetic Algorithm for Multiobjective Optimization: Formulation, Discussion and Generalization", 5th International Conference Genetic Algorithm, 416-423, 1993
[8] A.Y. Zomaya and Y.H. Teh, "Observation on Using Genetic Algorithm for Dynamic Load-Balancing", IEEE Transaction on Parallel and Distributed Systems, 12(9):899-911, September 2001
[9] Mihai Horia Zaharia, Florin Leon, Dan Galea, "Parallel Genetic Algorithm for Cluster Load Balancing", Proceesings 3th European Conference on Intelligent Systems and Technologies (ECIT 2004), Iasi, Romania, July 21-23, 2004
[10] Ramanujam Nedunchelian, Kalyanaraman Koushik, Nagappan Meiyappan, Viswanathan Raghu, "Dynamic Task Scheduling Using Parallel Genetic Algorithms For Heterogeneous Distributed Computing", Proceedings of the 2006 International Conference on Grid Computing & Applications (GCA) 2006 p. 82-88, Las Vegas, Nevada, USA, June 26-29, 2006
[11] Z. Akbar and L.T. Handoko, "Web-based Interface in Public Cluster", Proceeding of the 9th International Conference on Information Integration and Web-based Application and Services, Jakarta, Indonesia, 2007.